\documentclass[prd,nofootinbib,onecolumn,preprint,11pt]{revtex4-2}

\usepackage{graphicx}
\usepackage{bm}
\usepackage{float}
\usepackage{subfigure}
\usepackage{amsmath}
\usepackage{amsfonts}
\usepackage{color}
\usepackage{slashed}
\usepackage{fancyhdr}
\usepackage[colorlinks=true,linkcolor=black,citecolor=black,filecolor=black,urlcolor=black,unicode]{hyperref}

\newcommand{\vph}{\vphantom{\dot A}}

\newcommand{\GN}{G_{\rm N}}

\newcommand{\dd}{\text{d}}

\begin{document}

\title{Cherenkov Gravitational Radiation During the Radiation Era}

\author{Yi-Zen Chu$^{1,2}$ and Yen-Wei Liu$^3$}
\affiliation{
$\,^1$Department of Physics, National Central University, Chungli 32001, Taiwan \\
$\,^2$Center for High Energy and High Field Physics (CHiP), National Central University, Chungli 32001, Taiwan \\
$\,^3$Department of Physics, National Tsing-Hua University, Hsinchu 30013, Taiwan
}

\begin{abstract}
\noindent Cherenkov radiation may occur whenever the source is moving faster than the waves it generates. In a radiation dominated universe, with equation-of-state $w = 1/3$, we have recently shown that the Bardeen scalar-metric perturbations contribute to the linearized Weyl tensor in such a manner that its wavefront propagates at acoustic speed $\sqrt{w}=1/\sqrt{3}$. In this work, we explicitly compute the shape of the Bardeen Cherenkov cone and wedge generated respectively by a supersonic point mass (approximating a primordial black hole) and a straight Nambu-Goto wire (approximating a cosmic string) moving perpendicular to its length. When the black hole or cosmic string is moving at ultra-relativistic speeds, we also calculate explicitly the sudden surge of scalar-metric induced tidal forces on a pair of test particles due to the passing Cherenkov shock wave. These forces can stretch or compress, depending on the orientation of the masses relative to the shock front's normal.
\end{abstract}

\maketitle

\newpage

\section{Motivation}
\label{Section_Introduction}

We have recently, in \cite{Chu:2020sdn,Chu:2019ndv,Chu:2016ngc}, studied the gravitational perturbations $\chi_{\mu\nu}$ of the background geometry of a 4D universe driven by a perfect fluid with equation-of-state $w = 1/3$ (for instance, one comprised of relativistic neutrinos and photons) -- namely\footnote{Some words on notation: Our spacetime indices are Greek, running over $0$ (time) and $1$ through $3$ (space); while spatial ones are Latin/English. Einstein summation convention is in force unless otherwise stated. Unit spatial vectors have length $1$ in Euclidean space; and dot products are carried out with its metric $\delta_{ij}$.}
\begin{align}
	\label{RadiationDominatedUniverse}
	\dd s^2 &= a[\eta]^2 \left( -\dd\eta^2 + \dd\vec{x}\cdot\dd\vec{x} + \chi_{\mu\nu} \dd x^\mu \dd x^\nu \right), \\
	a[\eta] &= \eta/\eta_0 , \qquad
	x^\mu \equiv \left( \eta,\vec{x} \right)
\end{align}
-- sourced by an isolated hypothetical astrophysical system with energy-momentum-shear-stress tensor $\,^{\text{(a)}}T_{\mu\nu}$. Specifically, we solved within a gauge-invariant formalism the linearized Einstein's equations
\begin{align}
\delta_1 G_{\mu\nu} - 8 \pi \GN \delta_1 {}^{\text{(f)}}T_{\mu\nu}  &= 8\pi \GN {}^{\text{(a)}}T_{\mu\nu} ,
\end{align}
where $\delta_1 G_{\mu\nu}$ and $\delta_1 {}^{\text{(f)}}T_{\mu\nu}$ are, respectively, the pieces of Einstein's tensor $G_{\mu\nu}$ and the background fluid's stress-energy tensor ${}^{\text{(f)}}T_{\mu\nu}$ with precisely one power of $\chi_{\alpha\beta}$; and we have discarded all the $\chi_{\alpha\beta}$ dependence in the stress tensor $\,^{\text{(a)}}T_{\mu\nu}$. The key result we obtained was the following. The portions of $\chi_{\alpha\beta}$ that simultaneously transformed as scalars and remained invariant under infinitesimal coordinate transformations, otherwise dubbed by cosmologists as the Bardeen scalars $\Phi$ and $\Psi$, contribute to the dynamics of the linearized Weyl tensor through the expression
{\allowdisplaybreaks\begin{align}
	\delta_1  C^{(\Psi)i}{}_{0j0}[\eta,\vec x]  = \,& -2 G_\text N \int_{\mathbb{R}^{3}} \dd^{3} \vec x' \left(\delta_{ij}-3\widehat R_i\widehat R_j\right)  \nonumber\\
	&\times \Bigg\{ \frac1{R\eta^2} \int_0^\infty \dd \eta' \, \delta \left[ \eta - \eta' - \sqrt 3 R \right]  \left( {}^{(\text{a})}  T_{00}\big[\eta',\vec x'\big] + {}^{(\text{a})}  T_{ll} \big[\eta',\vec x' \big]  \right)  \notag\\
	&\qquad\qquad + \frac1{3R^3}\int_0^{\eta - \sqrt 3 R - 0^+}  \dd\eta' \, \frac{ \eta^3 - \eta'^3 }{ \eta^3 \eta'}  \left( {}^{(\text{a})}  T_{00}[\eta',\vec x'] + {}^{(\text{a})}  T_{ll}[\eta',\vec x']  \right)  \Bigg\} ,
	\label{Weyl_ScalarContribution_4DRadiation}
\end{align}}%
with $\delta[\dots]$ denoting Dirac's delta function; ${}^{(\text{a})} T_{ll} \equiv \delta_{mn} {}^{(\text{a})}  T_{mn}$ denotes the spatial trace of the matter stress-energy tensor; $R \equiv |\vec{x}-\vec{x}'|$ is the Euclidean coordinate distance between some observer at $\vec{x}$ and source at $\vec{x}'$; and $\widehat{R}^i = (x^i-x'^i)/R$ is the associated unit vector.

From the retarded time $\eta' = \eta-\sqrt{3}R$ in the second line of eq. \eqref{Weyl_ScalarContribution_4DRadiation}, we see that these scalar-metric perturbations induced traceless-tidal-forces propagate at the speed of sound $\sqrt{w} = 1/\sqrt{3}$. Additionally, the final line tells us the same waves also permeate the interior of the acoustic cone; i.e., they develop tails.

What happens to the Bardeen-scalars portion of the linearized Weyl tensor in eq. \eqref{Weyl_ScalarContribution_4DRadiation} when the source associated with ${}^{\text{(a)}}T_{\mu\nu}$ moves through the background fluid faster than the sound speed $\sqrt{w}=1/\sqrt{3}$? This situation is analogous to an electrically charged particle moving through a medium at a speed $v$ greater than the latter's effective speed of light $c_\text{eff} < 1$, in units where the vacuum light speed is unity. Whenever $c_\text{eff} < v < 1$, the electric charge will in fact outrun the electromagnetic signal it engenders. A Cherenkov shock front develops that divides space into two distinct regions: one where every point in space is causally linked to up to two retarded locations of the charge; and the other where the charge's worldline lies completely outside of the effective past null cone of every point in space. As we shall see in this paper, a similar scenario plays out for the gravitational case at hand.

In \S \eqref{Section_Cherenkov} we will lay out the generalities behind Cherenkov radiation emitted in a background conformally-flat geometry, such as the cosmological spacetime we are interested in. Following that, \S \eqref{Section_BH} and \S \eqref{Section_String} will specialize respectively to a hypothetical primordial black hole and straight cosmic string moving at supersonic speeds ($\sqrt{w} < v < 1$); and we will work out their Cherenkov traceless-tidal-forces signatures in the simple context of linear motion at ultra-relativistic speeds. Finally, in \S \eqref{Section_Summary}, we will discuss our results and point out potential future directions. In appendix \S \eqref{Section_Integral} we evaluate a key integral encountered in \S \eqref{Section_String}.

\section{Cherenkov Radiation: Generalities}
\label{Section_Cherenkov}

In this section, we shall elaborate on the key observation that Cherenkov radiation is expected to form whenever its source is moving faster than the waves themselves. As we shall see: the shape of the Cherenkov shock front, which sharply divides space into one region without any signal whatsoever and one with non-trivial signals, is completely determined by causality considerations. 

{\bf Spacetime Perspective} \qquad We begin by noting that the wave front encoded within the Dirac $\delta$-function term of eq. \eqref{Weyl_ScalarContribution_4DRadiation} is the same wave front that any massless wave would exhibit in the fictitious Minkowski spacetime
\begin{align}
\label{FakeMinkowski}
-\dd \eta^2 + \frac{\dd\vec{x}\cdot\dd\vec{x}}{w} = -\dd \eta^2 + \dd\vec{y}\cdot\dd\vec{y} ,
\end{align}
with $\vec{y} \equiv \vec{x}/\sqrt{w}$. From this perspective, a strictly supersonic source sweeps out a worldline (or world-tube) whose tangent vector(s) $\dd Y^\mu/\dd \lambda$ is {\it spacelike} with respect to the right hand side of eq. \eqref{FakeMinkowski} everywhere along its spacetime trajectory. That is, $|\dd\vec{Y}/\dd\eta| > 1$; or, in terms of the $(\eta,\vec{x})$ coordinate system, $|\dd \vec{X}/\dd \eta| > \sqrt{w}$. Whereas a strictly subsonic source is {\it timelike} with respect to the right hand side of eq. \eqref{FakeMinkowski}. 

\begin{figure}[H]
	\begin{center}
		\includegraphics[width=3in]{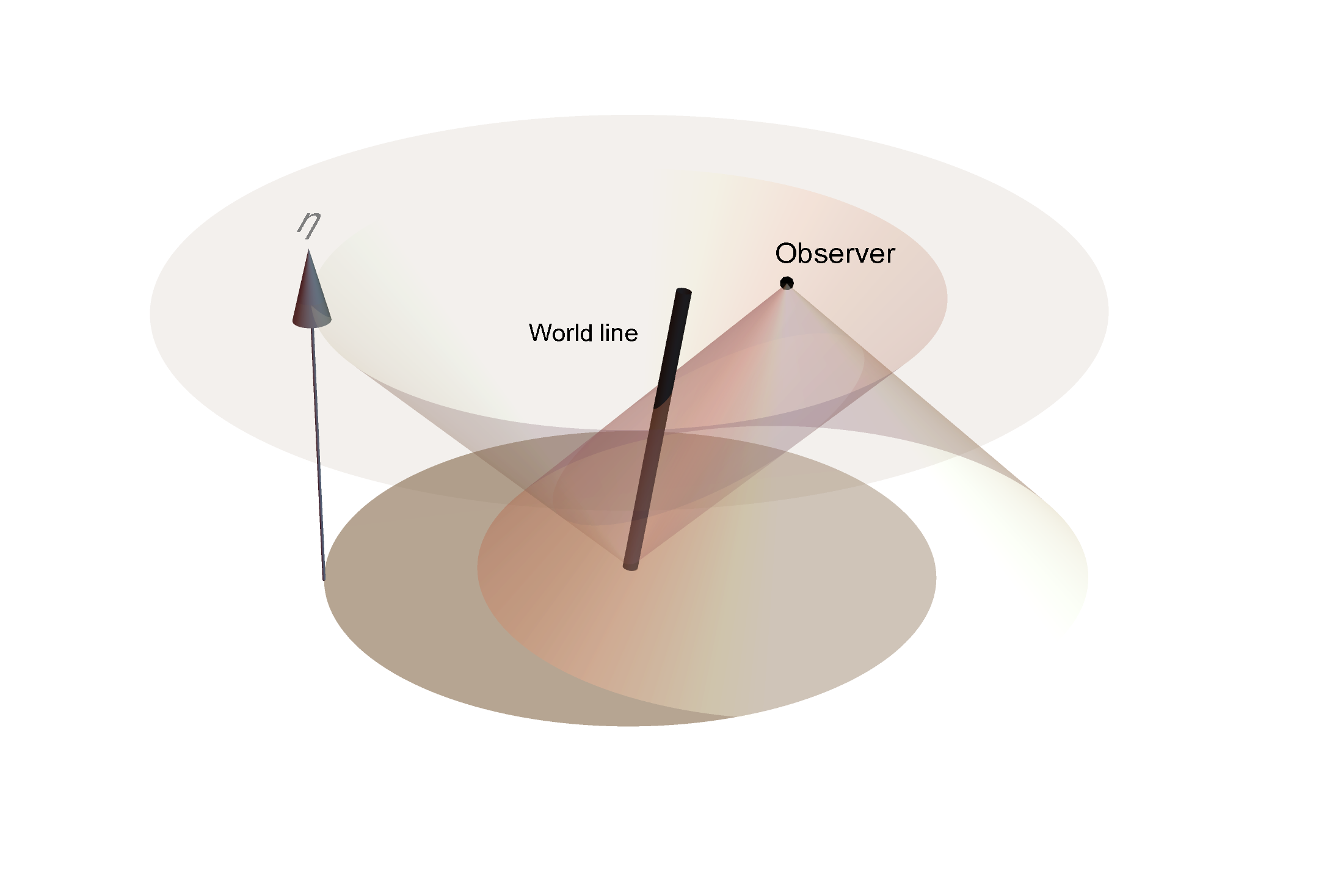}
		\includegraphics[width=3in]{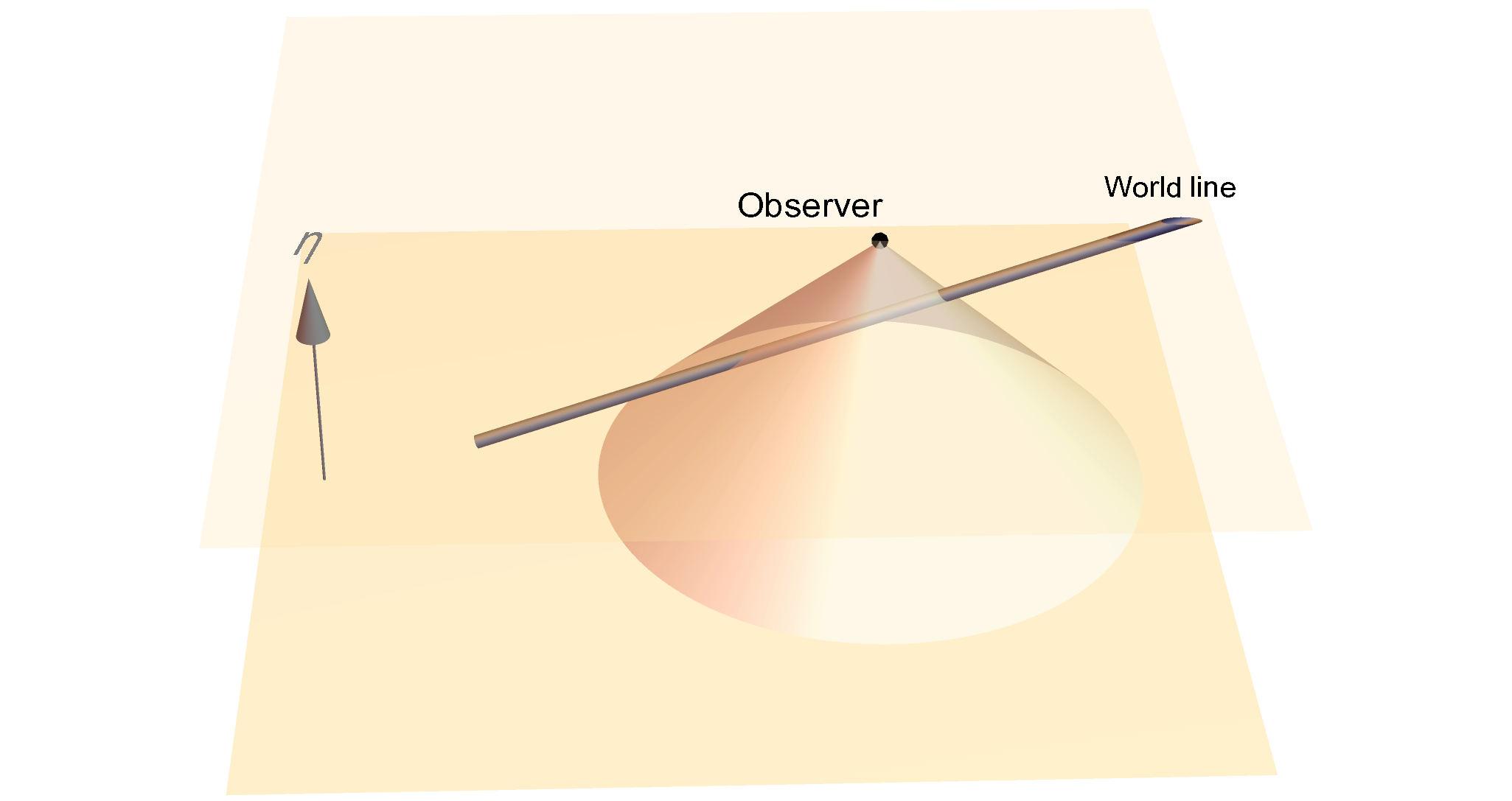} \\
		\includegraphics[width=4in]{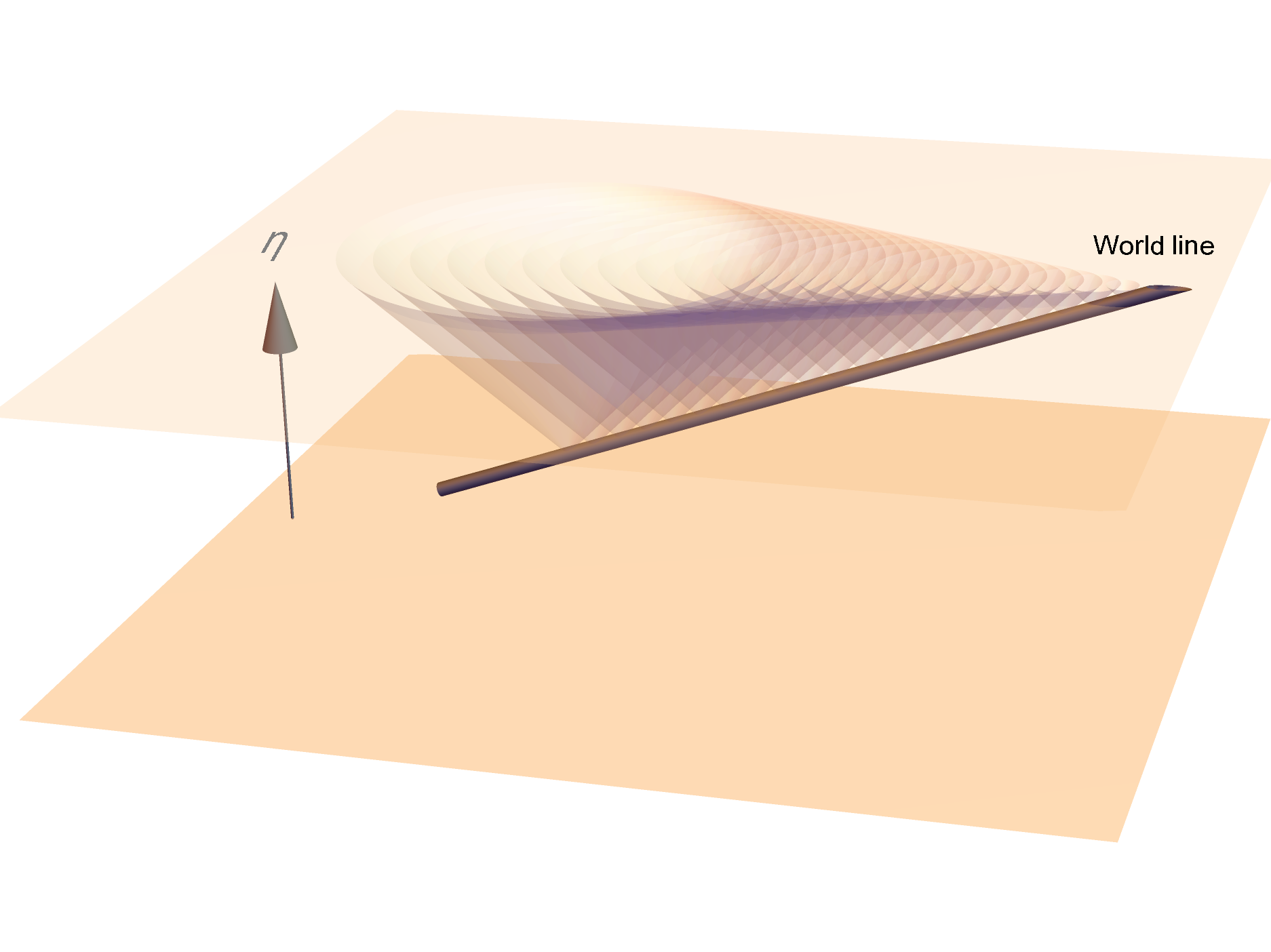}
		\caption{{\it Cherenkov radiation: spacetime perspective.} In all the above figures, we employ the $(\eta,\vec{y})$ coordinate system on the right hand side of eq. \eqref{FakeMinkowski}; with time running upwards. On the top left panel, we illustrate the fact that, for an observer on some constant $\eta > 0$ surface (top disk), as long as she is situated within the acoustic particle horizon of the source -- whose world tube is the black thick line -- there will always be a non-trivial signal because her past null cone will necessarily intersect with it. The dark disk lies on the $\eta=0^+$ surface infinitesimally near the Big Bang, and its radius corresponds to the size of the horizon: it is the same radius of the intersection region between the top constant $\eta$ surface with the forward light cone of the source at $\eta \to 0^+$. On the top right panel, we turn to consider instead a spacelike source. We see that such a world tube would generically intersect the observer's past null cone at two locations, corresponding to two `retarded times'. Only when the world tube is tangent to the null cone does the two reduce to one; the observer is then located on the Cherenkov cone which clearly divides space into a signal-free and a non-zero signal region. To see this, we move on to the bottom panel, where we have the same spacelike source's world tube, but attach forward null cones on various points along it. The Cherenkov shock front for a fixed $\eta > 0$ surface is the envelope of the locus of all such null cones emanating from the world tube.}
		\label{CherenkovFigure_SpacetimePerspective}
	\end{center}
\end{figure}

These in turn imply, as long as an observer at $(\eta,\vec{y})$ lies within the acoustic particle horizon of the strictly subsonic source, it will always receive a wavefront signal from the retarded location(s) of the source; namely, from the intersection of the latter's world tube with the past null cone of the observer. On the other hand, for a strictly supersonic source and a fixed observer time $\eta$, it may be possible to find regions of space that are causally disconnected from the source. Whenever the backward null cone of some observer at $(\eta,\vec{y})$ does intersect the supersonic spacetime trajectory, however, we see that -- because the source trajectory necessarily needs to enter and exit the null cone -- there must be exactly two retarded locations from which the observer receives the wavefront signals. The only exception to this statement occurs when the spacetime trajectory of the source ends at the Big Bang {\it inside} the backward null cone of the observer, hence leaving only one retarded location. Finally, consider the scenario where the worldline of a supersonic spatial-point source lies {\it tangent} to the backward light cone of $(\eta,\vec{y})$. For a fixed $\eta$, the locus of all such $\vec{y}$ cleanly demarcates two spatial regions, one causally connected and the other disconnected from the source.

We illustrate some of these statements in Fig. \eqref{CherenkovFigure_SpacetimePerspective}.

{\bf Huygens' Principle Perspective} \qquad Alternatively, the causal structure of massless signals may also be understood through Huygens' principle. For technical simplicity, in this work we shall focus on a spatial point or string source executing linear motion along the $1-$axis.

For the point source, because the setup is invariant under rotations along the $1-$axis, we may focus on the 2D cross section of 3D space that contains it, which we will name the $(y^1,y^2)$ plane. By invariance under parity $y^2 \to -y^2$, we may further focus only on the $y^2 > 0$ sector. Now, suppose our source passes through the spacetime point $Y^\mu[\eta_r] \equiv (\eta_r,\vec{Y}[\eta_r])$. Then, Huygens' principle says the spatial wavefront at a given instant $\eta > \eta_r$ due to the source at $Y^\mu[\eta_r]$ is the infinitesimally thin spherical shell of radius $\eta-\eta_r$ centered at but moving away from $\vec{Y}[\eta_r]$ at unit speed with respect to the $(\eta,\vec{y})$ system in eq. \eqref{FakeMinkowski}. When the point source is moving at subsonic speed $v < 1$, these wavefronts will outrun it. But when the source is supersonic, $v > 1$, the wavefronts evaluated on the $1-$axis will pile up on top of it because they move slower than $v$: the point source is thus the apex of the Cherenkov shock wave. Moreover, for a fixed $\eta$, the spatial Cherenkov front itself must be the envelope of all the ``Huygens' shells" emanating from retarded locations of the point source along the $1-$axis. Every spatial location $\vec{y}$ on the Cherenkov front must therefore lie tangent to a single Huygens' shell centered at $\vec{Y}[\eta_r]$, corresponding to the single point of intersection between the backward light cone of $(\eta,\vec{y})$ with the source's spacetime trajectory $Y^\mu$ -- as we have already argued above when discussing the spatial boundary of causal connectedness. Since this $\vec{y}$ corresponds to a unique $\eta_r$, we may view it as a function of this retarded time; i.e., $\vec{y}=\vec{y}[\eta_r]$. Altogether, causality tells us the Cherenkov front obeys
\begin{align}
\eta - \eta_r = |\vec{y}[\eta_r] - \vec{Y}[\eta_r]| .
\end{align}
Taking the derivative with respect to retarded time,
\begin{align}
- 1 = \widehat{R} \cdot \left(\dot{\vec{y}} - \dot{\vec{Y}}\right) ,
\end{align}
where $\widehat{R} \equiv (\vec{y}-\vec{Y})/|\vec{y}-\vec{Y}|$ and the overdot denotes a $\eta_r$ derivative. Since $\dot{\vec{y}}$ is tangent to the Cherenkov front, and since $\widehat{R}$ is the radial vector pointing from the retarded location of the point source to the location $\vec{y}$ on the Cherenkov front, we must have $\widehat{R} \cdot \dot{\vec{y}} = 0$. Therefore,
\begin{align}
\label{CherenkovFront_Cosine}
\cos \vartheta 
\equiv \widehat{e}_1 \cdot \widehat{R}[\eta_r] 
= \frac{\sqrt{w}}{|\dot{\vec{X}}[\eta_r]|} .
\end{align} 
Here, we have denoted the unit vector parallel to the $1-$axis as $\widehat{e}_1$; and reverted back to the $(\eta,\vec{x})$ system in eq. \eqref{FakeMinkowski} by setting $X^\mu[\eta_r] \equiv (\eta_r,\sqrt{w}\vec{Y}[\eta_r])$; i.e., $\vec{X}$ is the spatial location of the point source within the geometry of the left hand side of eq. \eqref{FakeMinkowski}. Notice from eq. \eqref{CherenkovFront_Cosine}, if the speed of the point source approaches the subsonic region from above, $|\dot{\vec{X}}| \to \sqrt{w}+0^+$, the angular spread of the Cherenkov front's tip will grow wider until it becomes flush against the 2D plane orthogonal to $\widehat{e}_1$.

\begin{figure}[!ht]
	\begin{center}
		\includegraphics[width=3in]{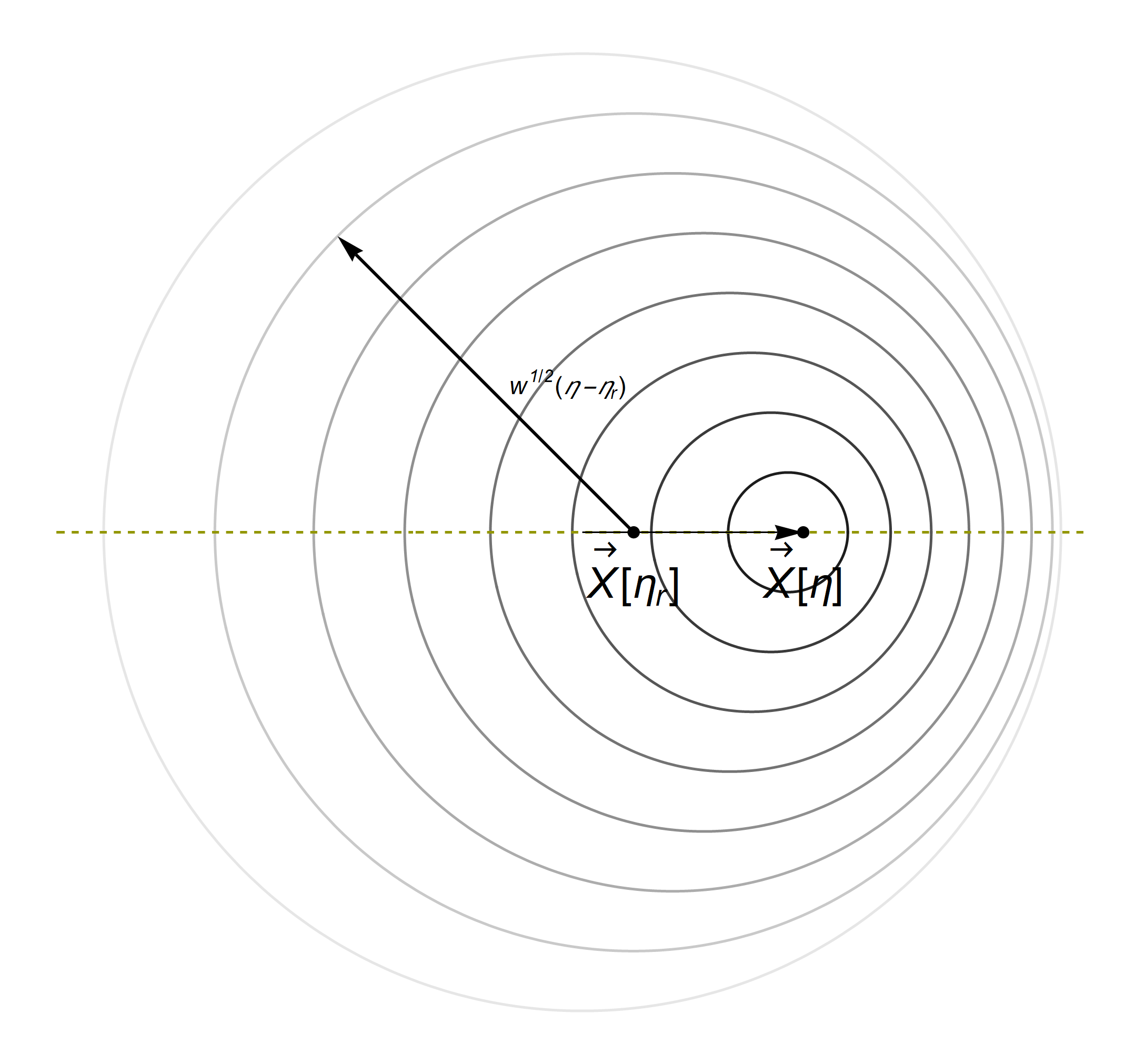}
		\includegraphics[width=3.5in]{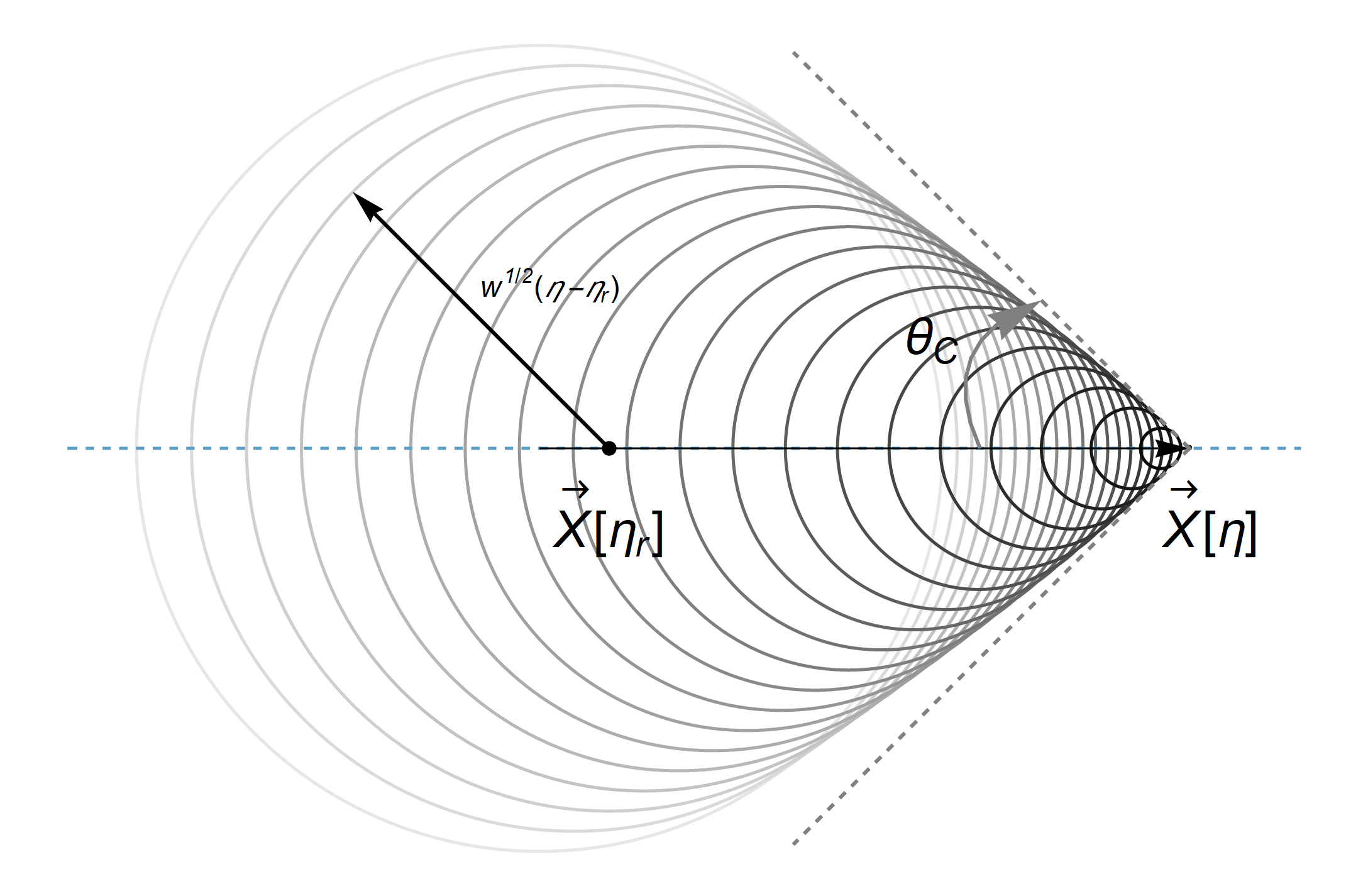}
		\caption{{\it Cherenkov radiation: Huygens' perspective.} These plots show, for some fixed time $\eta$, the retarded spatial locations of the point source (points on the two horizontal black arrows) moving along the $\widehat{e}_1$ axis (dotted horizontal line) strictly towards the right. The current position of the point is the arrow's tip. Each circle is the wavefront that emanated from the point $\vec{X}[\eta_r]$ on the horizontal black arrow lying on the former's center, with radius equal to $\sqrt{w}=1/\sqrt{3}$ times the time elapsed $\eta-\eta_r$, where $\eta_r$ is the retarded time of that location. Since the further left on the arrow means farther to the past in retarded time, this is why their associated wavefront circles are larger than those centered near the arrow's tip. {\it Left panel:} When the point source moves slower than $\sqrt{w}$, basic kinematics indicate it is always outrun by the wavefronts evaluated on the $\widehat{e}_1$ to its right. {\it Right panel:} When the point source moves faster than $\sqrt{w}$, however, no wavefront to its right can outrun it. What occurs instead is the wavefronts begin to pile up; the envelope of these circles define the Cherenkov shock wave -- this is the constant$-\eta$ snapshot of the bottom panel in Fig. \eqref{CherenkovFigure_SpacetimePerspective} -- outside of it, no signal can be detected.}
		\label{CherenkovFigure_HuygensPerspective}
	\end{center}
\end{figure}

The {\it acute} angle made by the Cherenkov cone with the axis of motion $\widehat{e}_1$ is $(\pi/2)-\vartheta$ when $\dot{\vec{X}}$ is evaluated at its current values -- by taking the limit $\eta_r \to \eta$ of eq. \eqref{CherenkovFront_Cosine},
\begin{align}
\label{CherenkovFront_Angle}
\cos \theta_\text{C} 
= \lim_{\eta_r \to \eta} \cos\left[ \vartheta - \frac{\pi}{2} \right] 
= \sqrt{1-\frac{w}{\dot{\vec{X}}^2[\eta]}} . 
\end{align}
In the ultra-relativistic limit, where $|\dot{\vec{X}}[\eta]| \to 1$, we have
\begin{align}
	\label{CherenkovFront_Angle_UR}
	\cos \theta_\text{C} 
	&\to \sqrt{1-w} = \sqrt{\frac{2}{3}} .
\end{align}
Now that we have the cosine of the angle between the $1-$axis and the line joining the spatial location $\vec{x}[\eta_r]$ on the Cherenkov front and the retarded location of the point source responsible for it, Euclidean 3D vector calculus then tells us 
\begin{align}
\label{CherenkovFront_PointMass_Step1}
\vec{x}_\text{C}[\eta_r] 
= \vec{X}[\eta_r] + \widehat{e}_1 \frac{w(\eta-\eta_r)}{|\dot{\vec{X}}[\eta_r]|}
+ \widehat{e}_2 \sqrt{w} (\eta-\eta_r) \sqrt{1-\frac{w}{\dot{\vec{X}}^2[\eta_r]}} .
\end{align}
The full spatial Cherenkov front -- really a cone -- at a given $\eta$ is simply the 2D surface of revolution obtained by rotating the locus of all points consistent with eq. \eqref{CherenkovFront_PointMass_Step1}. Let $\widehat{\rho}$ be the unit radial vector perpendicular to $\widehat{e}_1$, lying on the $(x^2,x^3)$ plane so that $\widehat{\rho} \cdot \widehat{e}_1 = 0$. Then our 2D Cherenkov cone, due to the point source at a given time $\eta$, may now be paramterized by $\eta_r$ and $\widehat{\rho}$ as
\begin{align}
	\label{CherenkovFront_PointMass}
	\vec{x}_\text{C}[\eta;\eta_r,\widehat{\rho}] 
	= \vec{X}[\eta_r] + \widehat{e}_1 \frac{w(\eta-\eta_r)}{|\dot{\vec{X}}[\eta_r]|}
	+ \widehat{\rho} \sqrt{w} (\eta-\eta_r) \sqrt{1-\frac{w}{\dot{\vec{X}}^2[\eta_r]}} .
\end{align}

If our source were instead an infinite straight string with zero thickness, aligned parallel to the $3-$axis but still moving along the $1-$axis, then the considerations are very similar to the point source case due to translation symmetry along the $3-$direction. After all, a straight string in 3D is a point on the 2D cross section perpendicular to its length. That is, we only need to study the problem on the $(x^1,x^2)$ plane for arbitrary but fixed $x^3$. On this $(x^1,x^2)$ plane, the wavefront emanating from some retarded location $\vec{X}[\eta_r]$ is the same as that of the point source case above, because the signals from the rest of the string $X^3[\eta_r] \neq x^3$ has to travel further to get to the same point on the constant $x^3$ plane, and therefore contributes only to the tail and not to the wavefront. 
\begin{align}
	\label{CherenkovFront_String}
	\vec{x}_\text{C}[\eta;\eta_r,x^3,\pm] 
	= \vec{X}[\eta_r] + \widehat{e}_1 \frac{w(\eta-\eta_r)}{|\dot{\vec{X}}[\eta_r]|}
	\pm \widehat{e}_2 \sqrt{w} (\eta-\eta_r) \sqrt{1-\frac{w}{\dot{\vec{X}}^2[\eta_r]}} .
\end{align}
The primary difference with the point source case above lies in the full spatial Cherenkov front. Here, for the string case, it is really an infinite wedge at a given $\eta$, gotten as the 2D surface of translation by shifting eq. \eqref{CherenkovFront_String} along the positive and negative $3-$direction to $x^3 \to \pm \infty$. Note that the {\it acute} angle the Cherenkov wedge makes with the axis of motion $\widehat{e}_1$ is the same eq. \eqref{CherenkovFront_Angle} as the point source case; and hence reduces to eq. \eqref{CherenkovFront_Angle_UR} in the ultra-relativistic limit.

We illustrate some of these statements in Fig. \eqref{CherenkovFigure_HuygensPerspective}.

\section{Supersonic Primordial Black Holes}
\label{Section_BH}
It has been speculated that the hypothetical but increasingly plausible inflationary phase of our early universe may produce enough over densities such that black holes could form upon the onset of the radiation domination era, due to gravitational collapse. (See \cite{Carr:2020xqk} for a review.) It is conceivable that some of these primordial black holes may be found at supersonic speeds, perhaps due to interactions with other black holes.

In this section, we will consider a single point mass $m$ -- approximating a primordial black hole -- moving along a geodesic in $g_{\mu\nu}=a^2\eta_{\mu\nu}$, the background radiation dominated universe of eq. \eqref{RadiationDominatedUniverse}. Denoting its spatial trajectory as $\vec{X}[\eta]$, its geodesic Lagrangian is
\begin{align}
	L_g 
	&\equiv a[\eta] \sqrt{ 1 - \dot{ \vec X }^2 } .
\end{align}
Spatial translation symmetry of the background universe then yields a conserved spatial momentum $\vec{p}_0 = -\partial L_g/\partial \dot{\vec{X}}$. These may be used together to deduce that the point mass' physical velocity with respect to the rest frame of the background geometry is
\begin{align}
	\label{PointMass_Velocity}
	\dot{\vec{X}} \equiv \frac{\dd \vec{X}[\eta]}{\dd\eta} = \frac{\vec{p}_0}{\sqrt{\vec{p}_0^2 + a[\eta]^2}} ,
\end{align}
for constant $\vec{p}_0$. If $\vec{X}_\text{bb}$ denotes its spatial location infinitesimally close to the Big Bang at $\eta = 0^+$, eq. \eqref{PointMass_Velocity} may be integrated to hand us
\begin{align}
\label{PointMass_Trajectory}
\vec{X}[\eta] 
= \vec{X}_\text{bb} + \eta_0 \cdot \text{arcsinh}\left[ \frac{a[\eta]}{p_0} \right] \cdot \vec{p}_0 ,
\qquad\qquad
p_0 \equiv |\vec{p}_0| .
\end{align}
The stress-energy tensor of a point mass $m$ with trajectory $X^\mu$ in an arbitrary curved spacetime with coordinates $x^\mu$ and metric $g_{\mu\nu}$ is
\begin{align}
\,^{\text{(a)}}T^{\mu\nu}[x] 
= m \int \dd x^0 \frac{\delta^{(4)}\big[x-X[x^0]\big]}{\sqrt[4]{g[x]g[X]}} \frac{1}{\sqrt{ -g_{\alpha\beta} \frac{\dd X^\alpha}{\dd x^0} \frac{\dd X^\beta}{\dd x^0} }} \frac{\dd X^\mu}{\dd x^0} \frac{\dd X^\nu}{\dd x^0}  .
\end{align}
Incorporating the geodesics of equations \eqref{PointMass_Velocity} and \eqref{PointMass_Trajectory}, our primordial black hole's stress tensor is
\begin{align}
	\label{PointMass_StressTensor}
	\,^{\text{(a)}}T_{\mu\nu}[\eta,\vec{x}]
	= \frac{m}{a[\eta]} \sqrt{1+\frac{\vec{p}_0^2}{a[\eta]^2}} 
	\left( \eta_{\mu 0} + \eta_{\mu i} \frac{p_0^i}{\sqrt{\vec{p}_0^2 + a[\eta]^2}} \right)
	\left( \eta_{\nu 0} + \eta_{\nu j} \frac{p_0^j}{\sqrt{\vec{p}_0^2 + a[\eta]^2}} \right)
	\delta^{(3)}[\vec{x}-\vec{X}[\eta]] .
\end{align}
{\bf Supersonic Speed} \qquad The supersonic condition is $|\dot{\vec{X}}| > \sqrt{w}$, which eq. \eqref{PointMass_Velocity} then informs us translates to the inequality
\begin{align}
	\frac{p_0}{a[\eta]} > \sqrt{\frac{w}{1-w}} = \frac{1}{\sqrt{2}} .
\end{align}
We shall assume this is obeyed in what follows. 

Firstly, the shape of the Cherenkov cone generated by our primordial black hole may be gotten by plugging equations \eqref{PointMass_Velocity} and \eqref{PointMass_Trajectory} into eq. \eqref{CherenkovFront_PointMass}; with the identifications 
\begin{align}
\widehat{e}_1 \leftrightarrow \vec{p}_0/ p_0 \equiv \widehat{p}_0
\qquad\qquad \text{ and } \qquad\qquad
\widehat{\rho} \cdot \vec{p}_0 = 0 .
\end{align}
The Cherenkov cone at $\eta$ produced by the supersonic point mass is
\begin{align}
\vec{x}_\text{C}[\eta;\eta_r,\widehat{\rho}]
&= \left( \eta_0 p_0 \cdot \text{arcsinh}\left[ \frac{a[\eta_r]}{ p_0 } \right]
+ w \left(\eta-\eta_r\right) \sqrt{1+\frac{a[\eta_r]^2}{\vec{p}_0^2}} \right) \widehat{p}_0 \nonumber\\
&\qquad\qquad\qquad\qquad
+ \sqrt{w} \left(\eta-\eta_r\right) \sqrt{1 - w \left( 1 + \frac{a[\eta_r]^2}{\vec{p}_0^2} \right)} \widehat{\rho} ,
\end{align}
where we have exploited the spatial translation symmetry of the background geometry to set $\vec{X}_\text{bb} = \vec{0}$, without any loss in generality. Recall, too, that $\eta_r$ has a 1-to-1 correspondence to the coordinate parallel to $\vec{p}_0$; and one angle is needed to describe the orientation of $\widehat{\rho}$ -- our 2D Cherenkov front is, as expected, parametrized by $2$ coordinates. 

{\bf Ultra-relativistic Limit} \qquad Next, we shall compute the contribution to the physical tidal forces from the above Bardeen-scalars Cherenkov radiation, as encoded within eq. \eqref{Weyl_ScalarContribution_4DRadiation}. To this end, we will specialize to the ultra-relativistic limit $\vec{p}_0^2 \gg a^2/2$, where the calculations are somewhat simpler. In this limit, the stress-energy tensor \eqref{PointMass_StressTensor} simplifies to 
\begin{align}
	\label{PointMass_StressTensor_Ultrarelativistic}
	{}^{\text{(a)}}T_{\mu\nu}[\eta,\vec{x}]
	= \frac{m p_0}{a[\eta]^2} 
	\left( \eta_{\mu 0} + \eta_{\mu i} \widehat p_0^i \right)
	\left( \eta_{\nu 0} + \eta_{\nu j} \widehat p_0^j \right)
	\delta^{(3)}[\vec{x}-\vec{X}[\eta]] .
\end{align}
The velocity in eq. \eqref{PointMass_Velocity} is now
\begin{align}
\label{PointMass_Velocity_UR}
\frac{\dd \vec{X}[\eta]}{\dd\eta} 
= \widehat{p}_0 \left( 1 + \mathcal{O}\left[a^2/\vec{p}_0^2\right] \right) ;
\end{align}
while the trajectory in eq. \eqref{PointMass_Trajectory} then reads
\begin{align}
\label{PointMass_Trajectory_UR}
\vec{X}[\eta] 
= \eta \cdot \widehat{p}_0 \left( 1 + \mathcal{O}\left[a^2/\vec{p}_0^2\right] \right) .
\end{align}

If we define $x_\parallel$ and $x_\perp$ to be, respectively, the observer's spatial coordinates along the $\vec{p}_0$ and $\widehat{\rho}$ directions, namely
\begin{align}
	x_\parallel &\equiv \vec{x} \cdot \widehat{p}_0, \\
	x_\perp		&\equiv \left\vert x^i \left( \delta^{ij} - \widehat{p}_0^i \widehat{p}_0^j \right) x^j \right\vert^{\frac{1}{2}} ,
\end{align}
this point mass generated Cherenkov cone may also be described by the constraint
\begin{align}
\label{PointMass_CherenkovCone}
x_\parallel - \eta 
= -x_\perp \cot \theta_\text{C} 
= -\sqrt{2} x_\perp .
\end{align}
For an observer at $(\eta,\vec{x})$, the retarded time(s) $\eta_r$ is the solution(s) to the acoustic cone condition
\begin{align}
\label{PointMass_RetardedCondition}
w (\eta-\eta_r)^2 = (\vec{x}-\vec{X}[\eta_r])^2 
\end{align}
subject to the constraint $\eta > \eta_r$. The ultra-relativistic trajectory in eq. \eqref{PointMass_Trajectory_UR} inserted into eq. \eqref{PointMass_RetardedCondition} produces a quadratic equation. This hands us
\begin{align}
	\eta - \eta_r^\pm = \frac{  x_\parallel - \eta \mp \sqrt{ w ( x_\parallel - \eta)^2 + ( w - 1 ) x_\perp^2 } }{w - 1} > 0 .
	\label{PointMass_RetardedTimes}
\end{align}
As expected, eq. \eqref{PointMass_RetardedTimes} yields two solutions.\footnote{Here and for the supersonic cosmic string case discussed in \S \eqref{Section_String}, we will only consider the processes late enough for the observer to receive two retarded signals after the Big Bang, namely, $\eta_r^\pm > 0$. For the ultra-relativistic point source, this amounts to the spacetime region restricted by $ \vec x^2 = x_\parallel^2 + x_\perp^2 > w \eta^2 $.}  For, we have already pointed out, from the perspective of the fictitious Minkowski spacetime of eq. \eqref{FakeMinkowski}, the supersonic point source is sweeping out a {\it spacelike} world line.

Because we are primarily interested in the features of the Cherenkov shock front itself, we now proceed to ignore the tail contribution from the third line of eq. \eqref{Weyl_ScalarContribution_4DRadiation} and label the $\delta-$function acoustic-cone piece `direct'. The 3D Dirac delta function of eq. \eqref{PointMass_StressTensor_Ultrarelativistic} is collapsed by the $\mathbb{R}^3$ spatial integral over $\vec{x}'$. Taking into account the shape of the Cherenkov cone in eq. \eqref{PointMass_CherenkovCone} and the stress tensor contributions from the two retarded locations in eq. \eqref{PointMass_RetardedTimes}, we find
\begin{align}
\delta_1  C^{(\Psi|\text{Direct})i}{}_{0j0}[\eta,\vec x] 
&= - 4 G_\text N m \frac{p_0}{ a[\eta]^2 } \Theta\left[ \eta - x_\parallel - \sqrt{2} x_\perp \right]  \nonumber\\
&\times \left( C_1[\eta,\vec x] \delta_{ij} -  C_2[\eta,\vec x] \widehat \rho^i \widehat \rho^j  - C_3[\eta,\vec x]  \left( \widehat \rho^i \widehat{p}_0^j +  \widehat \rho^j \widehat{p}_0^i \right) -  C_4[\eta ,\vec x] \widehat{p}_0^i \widehat{p}_0^j \right) ,
		\label{ScalarTidalForces_PP_Ultrarelativistic}
\end{align}
where $\Theta$ is the Heaviside step function; and $\eta_r^\pm$ are given in eq.~\eqref{PointMass_RetardedTimes} as
\begin{align}
	\eta_r^\pm & =  \frac{ 3 x_\parallel - \eta }{ 2 } \pm \frac{\sqrt 3}{ 2 } \sqrt{ ( x_\parallel - \eta)^2 - 2 x_\perp^2 }  ,
\end{align}
The scalar coefficients $C$'s are defined by the sum
\begin{align}
	C_i[\eta ,\vec x] = C_i^+[\eta ,\vec x] + C_i^-[\eta ,\vec x] ,
\end{align}
with the $C^+$ representing the contribution from the $\eta_r^+$ retarded location and $C^-$ from the $\eta_r^-$. They read 
{\allowdisplaybreaks
	\begin{align}
		C_1^\pm [\eta, \vec x ] & =   \frac{1}{  \sqrt{(x_\parallel-\eta )^2-2 x_\perp ^2}} \frac{1}{ \left( \eta_r^\pm \right)^2 }  , \\
		C_2^\pm [\eta, \vec x ] & =   \frac{ 1 }{  \sqrt{(x_\parallel-\eta )^2-2 x_\perp ^2}} \frac{ 9 x_\perp^2 }{ \left( \eta_r^\pm \right)^2 \left( \eta - \eta_r^\pm \right)^2 }   , \\
		C_3^\pm [\eta, \vec x ] & =   \frac{ 1 }{  \sqrt{(x_\parallel-\eta )^2-2 x_\perp ^2}} \frac{ 9 x_\perp \left( x_\parallel - \eta_r^\pm \right)  }{ \left( \eta_r^\pm \right)^2 \left( \eta - \eta_r^\pm \right)^2 }    , \\
		C_4^\pm [\eta, \vec x ] & =   \frac{ 1 }{  \sqrt{(x_\parallel-\eta )^2-2 x_\perp ^2}} \frac{ 9  \left( x_\parallel - \eta_r^\pm \right)^2  }{ \left( \eta_r^\pm \right)^2 \left( \eta - \eta_r^\pm \right)^2 } .
\end{align}}%
{\bf Tidal Forces Near the Cherenkov Cone} \qquad If the observer is close to the Cherenkov cone, we may parameterize her location by a small conformal distance $\ell$ away from the cone via
\begin{align}
	\ell = \left( \frac{\eta - x_\parallel}{\sqrt 2} - x_\perp  \right) \cos\theta_\text C = \sqrt{ \frac23 } \left( \frac{\eta - x_\parallel}{\sqrt 2} - x_\perp  \right)  ;
\end{align}
and assume $\ell / (\eta - x_\parallel) \ll 1$ and $ \ell / ( 3 x_\parallel - \eta) \ll 1$. The retarded times are now, to leading order,
\begin{align}
	\eta_r^\pm = \frac{ 3 x_\parallel - \eta }{2} - \frac{3^{\frac34} }{\sqrt{2}} \sqrt{\ell \left( \eta - x_\parallel \right) }  + \mathcal O \big[\ell ^\frac32\big] .
\end{align}
Then, the scalar coefficients $C_{1,2,3,4}$ can be expanded in powers of $\ell$ as follows:
{\allowdisplaybreaks
	\begin{align}
		C_1[\eta , \vec x] & =  \frac1{\sqrt{\ell }} \left( \frac{4 \sqrt{2}}{ 3^\frac14  ( 3 x_\parallel - \eta )^2 \sqrt{\eta - x_\parallel }}+\frac{\sqrt{2} \cdot 3^\frac14 \left(73 \eta ^2+81 x_\parallel^2-150 \eta x_\parallel \right) \ell }{( 3 x_\parallel - \eta )^4 (\eta - x_\parallel )^\frac32} + \mathcal O\left[\ell^2\right]  \right) , \\
		C_2[\eta, \vec x] & =   \frac1{\sqrt{\ell }} \left( \frac{8 \sqrt{2}}{ 3^\frac14  ( 3 x_\parallel - \eta )^2 \sqrt{\eta - x_\parallel } } + \frac{2 \sqrt{2} \cdot 3^\frac14  \left(105 \eta ^2+177 x_\parallel^2 - 278 \eta  x_\parallel \right) \ell }{( 3 x_\parallel - \eta )^4 (\eta - x_\parallel )^{\frac32}} + \mathcal O \left[ \ell^2 \right] \right) , \\
		C_3[\eta , \vec x] & =  \frac1{\sqrt{\ell }} \left( \frac{8}{ 3^\frac14  ( 3 x_\parallel - \eta )^2 \sqrt{\eta - x_\parallel } } + \frac{2 \cdot 3^\frac14  \left(45 \eta ^2-75 x_\parallel^2 - 14 \eta x_\parallel \right) \ell }{( 3 x_\parallel - \eta )^4 (\eta - x_\parallel )^{\frac32}} + \mathcal O \left[\ell^2 \right] \right) , \\
		C_4[\eta , \vec x] & = \frac1{\sqrt{\ell }} \left( \frac{4 \sqrt{2}}{ 3^\frac14  ( 3 x_\parallel - \eta )^2 \sqrt{\eta  - x_\parallel } } + \frac{\sqrt{2} \cdot 3^\frac14 \left(9 \eta ^2-111 x_\parallel^2 + 106 \eta x_\parallel \right) \ell }{( 3 x_\parallel - \eta )^4 (\eta - x_\parallel )^{\frac32}} + \mathcal O\left[ \ell^2 \right] \right) .
\end{align}}%
To leading order in the $\ell$ expansion, the Bardeen-scalar induced tidal forces read
\begin{align}
\label{Weyl_BH_UC_NearCone}
\hspace{-0.5cm} \delta_1  C^{(\Psi|\text{Direct})i}{}_{0j0}[\eta,\vec x] 
&= -  \frac{ 16 \sqrt2 G_\text N m  p_0}{ 3^\frac14 a[\eta]^2  (3x_\parallel - \eta )^2  \sqrt{ \eta - x_\parallel }  } \frac1{ \sqrt \ell } \Theta\left[ \eta - x_\parallel - \sqrt{2} x_\perp \right] \left(  \delta_{ij} -  3 \widehat u^i \widehat u^j \right) + \mathcal O \big[ \ell^\frac12 \big] ,
\end{align}
where the unit spatial vector $\widehat{u}$ points to the region of no signal and is perpendicular to the Cherenkov cone:
\begin{align}
	\widehat u &\equiv \sqrt{\frac 23} \widehat \rho + \frac1{\sqrt 3} \widehat{p}_0 . 
	\label{UnitNormal_BH}
\end{align}

As we can see from eq.~\eqref{Weyl_BH_UC_NearCone}, these scalar-induced tidal forces may in fact be greatly amplified in the proximity of the Cherenkov cone formed by a supersonic point mass, compensating for their relatively weak signals compared to their spin-2 tensor counterparts in relativistic fluid-driven cosmologies; the former had been estimated in \cite{Chu:2020sdn} to be Hubble-suppressed relative to the latter. This setup, thus, provides a scenario in which the detectability of such scalar tidal effects could potentially be enhanced.

\section{Supersonic Cosmic Strings}
\label{Section_String}
If cosmic strings formed due to some phase transition in the early universe (see, e.g., the review \cite{Vachaspati:2015}), their dynamics during the radiation era could generate not only spin-2 (tensor) gravitational waves but also the Bardeen-scalar ones we are currently studying. In this section, we turn to consider the scenario where an infinite straight cosmic string is moving perpendicular to its length, at supersonic speeds. As we shall witness, just like its primordial black hole counterpart, a Cherenkov shock front will develop. 

We will approximate the cosmic string as an infinitesimally thin relativistic wire with tension $\mu$, endowed with spacetime coordinates $\{ Z^\mu \}$ and intrinsic coordinates $\{ \xi^\text{A} \equiv (\tau,\sigma) | \text{A}=0,1 \}$. The induced metric on the string's world sheet is
\begin{align}
	h_\text{AB}[\xi] &= g_{\mu\nu}\big[ Z[\xi] \big] \frac{\partial Z^\mu}{\partial \xi^\text{A}} \frac{\partial Z^\nu}{\partial \xi^\text{B}} , 
\end{align}
and $\sqrt{|h|}$ is the square root of the absolute value of its determinant. The spacetime dynamics of the string itself is encoded by the area swept out by its worldsheet -- otherwise known as the Nambu-Goto action --
\begin{align}
	S_\text{NG} &= - \mu \int \dd^2 \xi \sqrt{|h|} .
	\label{NGAction}
\end{align} 
The equations of motion derived from eq.~\eqref{NGAction} are 
\begin{align}
	\frac1{ \sqrt{|h|} }  \frac{\partial}{\partial \xi^\text{A}} \bigg(  \sqrt{|h|} \, h^\text{AB}   \frac{\partial Z^\rho }{\partial \xi^\text{B}}  \bigg)  = - h^\text{AB} \frac{\partial Z^\mu}{\partial \xi^\text{A}} \frac{\partial Z^\nu}{\partial \xi^\text{B}}  \Gamma^\rho_{\mu\nu} , 
	\label{EoM_String}
\end{align}
and the corresponding stress-energy tensor is 
\begin{align}
	{}^{\text{(a)}}T^{\mu\nu}[x] = - \mu  \int \dd^2 \xi \sqrt{|h|} \,  \frac{\delta^{(4)} \big[ x - Z[\xi] \big]}{ \sqrt[4]{g[x] g[Z]}  }  h^\text{AB}[\xi] \frac{\partial Z^\mu}{\partial \xi^\text{A}} \frac{\partial Z^\nu}{\partial \xi^\text{B}} .
	\label{StressEnergyTensor_String}
\end{align}
{\bf Gauge Choice} \qquad In the background geometry of eq. \eqref{RadiationDominatedUniverse}, i.e., $g_{\mu\nu}=a^2\eta_{\mu\nu}$, we shall choose the worldsheet time coordinate to be\footnote{The classical dynamics of the Nambu-Goto string in an expanding universe has been studied, for example, by Turok and Bhattacharjee \cite{Turok:1984db} and de Vega, Larsen, and Sanchez \cite{deVega:1994yz}; a pedagogical discussion and review can also be found in Sec.~6.3 of the book by Vilenkin and Shellard \cite{Vilenkin:2000jqa}.}
\begin{align}
	\tau = \eta = Z^0 ;
\end{align}
and its spatial coordinate to be orthogonal to $\tau$, 
\begin{align}
	g_{\mu\nu} \dot{Z}^\mu (Z^\nu)' = 0 ;
\end{align}
where an overdot is a derivative with respect to $\eta$ and prime $\xi^1=\sigma$. The induced metric then takes the expression
\begin{align}
	\label{EoM_InducedMetric}
	h_\text{AB} = a^2 \cdot \text{diag}\left[\dot{\vec{Z}}^2-1,\vec{Z}'^2\right].
\end{align}
These gauge conditions reduce eq. \eqref{EoM_String} to
\begin{align}
	\label{EoM_String_v2}
	\ddot{\vec{Z}} + 2 \frac{\dot{a}[\eta]}{a[\eta]} \left(1-\dot{\vec{Z}}^2\right) \dot{\vec{Z}} + \sqrt{\frac{1-\dot{\vec{Z}}^2}{\vec{Z}'^2}} \left( \sqrt{\frac{1-\dot{\vec{Z}}^2}{\vec{Z}'^2}} \vec{Z}' \right)' = 0 .
\end{align}
{\bf Infinite Moving Straight String} \qquad We now specialize to an infinite straight string moving perpendicularly to its length:
\begin{align}
	\label{EoM_InfiniteString_C1}
	\vec{Z} = \vec{b}[\eta] + \sigma \cdot \widehat{n} .
\end{align}
The $\widehat{n}$ is a unit vector pointing along the string's length for a fixed time $\eta$. The $\vec{b}$ describes the motion perpendicular to $\widehat{n}$, and thus satisfies
\begin{align}
	\label{EoM_InfiniteString_C2}
	\dot{\vec{b}} \cdot \widehat{n} = 0 .
\end{align}
The ansatz in eq. \eqref{EoM_InfiniteString_C1} together with the constraint in eq. \eqref{EoM_InfiniteString_C2} converts eq. \eqref{EoM_String_v2} into
\begin{align}
	\label{EoM_String_v3}
	\dot{\vec{v}} + 2 \frac{\dot{a}}{a} \left(1-\vec{v}^2\right) \vec{v} = 0 ,
	\qquad\qquad
	\vec{v} \equiv \dot{\vec{b}}.
\end{align}
Eq. \eqref{EoM_String_v3} may also be obtained by plugging eq. \eqref{EoM_InfiniteString_C1} and \eqref{EoM_InfiniteString_C2} into the Nambu-Goto action in eq. \eqref{NGAction}, to yield
\begin{align}
	S_\text{NG} &= -\mu \int \dd\eta \int\dd\sigma L_\text{NG}, \\
	L_\text{NG} &= a^2 \sqrt{1-\vec{v}^2} .
\end{align}
The advantage of starting with $L_\text{NG}$ is that it allows us to identify the conserved momentum $\vec{p}_0 = -\partial L_\text{NG}/\partial \vec{v}$ associated with spatial translation symmetry; which in turn tells us
\begin{align}
	\label{EoM_String_v4}
	\vec{v} = \frac{\vec{p}_0}{\sqrt{\vec{p}_0^2 + a[\eta]^4}} .
\end{align}
Since there is no preferred spatial origin, we shall assume $\vec{b}[\eta=0^+] = \vec{0}$ near the Big Bang. Integrating eq. \eqref{EoM_String_v4} then leads us to (cf. eq. \eqref{EoM_InfiniteString_C1})
\begin{align}
\vec{Z}[\eta] = \widehat{p}_0 \cdot \eta \,_2F_1\left[ \frac{1}{2}, \frac{1}{4}; \frac{4}{5}; -\frac{a[\eta]^4}{\vec{p}_0^2} \right] + \sigma \widehat{n}.
\end{align}
The stress tensor in eq. \eqref{StressEnergyTensor_String} now reads
\begin{align}
{}^{\text{(a)}}T_{\mu\nu}[\eta , \vec x]  
&=  \mu     \delta^{(2)} \big[ \vec x_\perp - \vec b_\perp[\eta]\big]  \nonumber\\
&\times \left\{  \sqrt{ 1 + \frac{\vec p_0^2 }{a[\eta]^4}  } \left( \eta_{\mu 0} +  \eta_{\mu i} \frac{ p_0^i }{ \sqrt{ \vec p_0^2 + a[\eta]^4 } }  \right) \left( \eta_{\nu 0} +  \eta_{\nu j}   \frac{ p_0^j }{ \sqrt{ \vec p_0^2 + a[\eta]^4 } }   \right) -    \frac{\eta_{\mu i} \eta_{\nu j} \widehat n^i  \widehat n^j  }{ \sqrt{ 1 + \frac{ \vec p_0^2 }{a [\eta]^4 } } }  \right\} .
	\label{StressEnergyTensor_String_4DRadiation}
\end{align}
The $\vec{x}_\perp$ and $\vec{b}_\perp$ refer to the components of $\vec{x}$ and $\vec{b}$ perpendicular to $\widehat{n}$; namely,
\begin{align}
x_\perp^i &\equiv (\delta^{ij} - \widehat{n}^i \widehat{n}^j) x^j, \\
b_\perp^i &\equiv (\delta^{ij} - \widehat{n}^i \widehat{n}^j) b^j.
\end{align}
{\bf Supersonic and Ultra-Relativistic Limits} \qquad If the string is moving at supersonic speeds, this corresponds to $|\vec{v}| > \sqrt{w} = 1/\sqrt{3}$. Eq. \eqref{EoM_String_v4} translates this inequality into
\begin{align}
p_0 > \sqrt{\frac{w}{1-w}} \, a[\eta]^2 = \frac{a[\eta]^2}{\sqrt 2} .
\end{align}

Again for technical simplicity, we shall proceed to work out the tidal forces arising from the Bardeen-scalars' Cherenkov radiation by taking the ultra-relativistic limit $\vec{p}_0^2 \gg a^4/2$. In such a limit, the string's stress tensor \eqref{StressEnergyTensor_String_4DRadiation} simplifies to
\begin{align}
	{}^{\text{(a)}}T_{\mu\nu}[\eta , \vec x]  
	&=  \frac{ \mu p_0 }{ a[\eta]^2 } \left( \eta_{\mu 0} +  \eta_{\mu i} \widehat p_0^i  \right) \left( \eta_{\nu 0} +  \eta_{\nu j}  \widehat p_0^j  \right) \delta^{(2)} \big[ \vec x_\perp - \vec b_\perp[\eta]\big]  .
	\label{StressEnergyTensor_String_Ultrarelativistic}
\end{align}
To parametrize the Cherenkov wedge produced by the infinite string, we first define the orthogonal coordinate system on the 2D plane perpendicular to the string:
\begin{align}
w^1 &\equiv \vec{x}_\perp \cdot \widehat{p}_0, \\
w^2 &\equiv \vec{x}_\perp \cdot (\widehat{n} \times \widehat{p}_0) .
\end{align}
The cosmic string's Cherenkov wedge in the ultra-relativistic limit (cf. eq. \eqref{CherenkovFront_Angle_UR}) is therefore given by
\begin{align}
	\sqrt{2} |w^2| = \eta - w^1 .
\end{align}
For the same reasons as the point mass case, we shall focus on the acoustic-cone portion of the linearized Weyl tensor in eq. \eqref{Weyl_ScalarContribution_4DRadiation}, now sourced by the string configuration in eq. \eqref{StressEnergyTensor_String_Ultrarelativistic}. A direct calculation yields
{\allowdisplaybreaks
\begin{align}
	\label{ScalarTidalForces_String_Relativistic}
&\delta_1  C^{(\Psi|\text{Direct})i}{}_{0j0}[\eta,\vec x] \nonumber\\
&= - \frac{ 8 G_\text N \mu }{\eta^2} \Theta\left[ \eta - w^1 - \sqrt{2} |w^2| \right] \int_0^\infty \dd \eta'  \, \frac{ \Theta\left[ \eta - \eta' - \sqrt 3 R_\perp[\eta'] \right]  }{\sqrt{ ( \eta - \eta' )^2 - 3 R_\perp[\eta']^2 } } \\
&\qquad\qquad\qquad\qquad
\times \left( \delta_{ij} - 3 \widehat{n}^i \widehat{n}^j
- 9 \frac{  R^i_\perp[\eta']  R^j_\perp[\eta'] - R_\perp[\eta']^2  \widehat{n}^i \widehat{n}^j   }{  (\eta - \eta' )^2 } \right) \frac{p_0 \eta_0^2 }{\eta'^2}  , \notag\\
&= - \frac{ 8 G_\text N \mu p_0}{a[\eta]^2} \Theta\left[ \eta - w^1 - \sqrt{2} |w^2| \right] \bigg\{ \left(  \delta_{ij} - 3 \widehat{n}^i \widehat{n}^j \right) I_1[\eta,\vec x_\perp] \nonumber\\
&\qquad\qquad - 9 \widehat{p}_0^i \widehat{p}_0^j \left( I_2[\eta,\vec x_\perp] - 2w^1 I_3[\eta,\vec x_\perp] + (w^1)^2 I_4[\eta,\vec x_\perp] \right)  \notag\\
&\qquad\qquad  - 9 \left( \widehat{p}_0^i (\widehat{n} \times \widehat{p}_0)^j + \widehat{p}_0^j (\widehat{n} \times \widehat{p}_0)^i \right) \left( w^1 w^2 I_4[\eta,\vec x_\perp] - w^2 I_3[\eta,\vec x_\perp]  \right) \nonumber\\
&\qquad\qquad - 9 (\widehat{n} \times \widehat{p}_0)^i (\widehat{n} \times \widehat{p}_0)^j \left( (w^2)^2 I_4[\eta,\vec x_\perp] \right)
+ 9 \widehat{n}^i \widehat{n}^j  \left( I_2[\eta,\vec x_\perp] - 2w^1 I_3[\eta,\vec x_\perp] + \vec{x}_\perp^2 I_4[\eta,\vec x_\perp] \right) \bigg\} , \nonumber
\end{align}}%
where 
\begin{align}
\vec R_\perp[\eta'] & = \vec{x}_\perp - \eta' \widehat{p}_0 , \\
R_\perp[\eta'] 		& = |\vec{R}_\perp[\eta']|, \label{2DConformalDistance}
\end{align}
and the scalar integrals involved are defined by
{\allowdisplaybreaks
\begin{align}
		\label{I1}
		I_1[\eta,\vec x_\perp] & \equiv \int_0^\infty \dd \eta'  \, \frac{ \Theta\left[ \eta - \eta' - \sqrt 3 R_\perp[\eta'] \right]  }{\sqrt{ ( \eta - \eta' )^2 - 3 R_\perp[\eta']^2 } }  \frac1{\eta'^2} , \\
		\label{I2}
		I_2[\eta,\vec x_\perp] & \equiv \int_0^\infty \dd \eta'  \, \frac{ \Theta\left[ \eta - \eta' - \sqrt 3 R_\perp[\eta'] \right]  }{\sqrt{ ( \eta - \eta' )^2 - 3 R_\perp[\eta']^2 } }  \frac1{ (\eta - \eta')^2}  , \\
		\label{I3}
		I_3[\eta,\vec x_\perp] & \equiv \int_0^\infty \dd \eta'  \, \frac{ \Theta\left[ \eta - \eta' - \sqrt 3 R_\perp[\eta'] \right]  }{\sqrt{ ( \eta - \eta' )^2 - 3 R_\perp[\eta']^2 } } \left( \frac{1}{\eta ^2 \eta'}-\frac{1}{\eta ^2 (\eta'-\eta )}+\frac{1}{\eta  (\eta'-\eta )^2} \right) ,  \\ 
		\label{I4}
		I_4[\eta,\vec x_\perp] & \equiv \int_0^\infty \dd \eta'  \, \frac{ \Theta\left[ \eta - \eta' - \sqrt 3 R_\perp[\eta'] \right]  }{\sqrt{ ( \eta - \eta' )^2 - 3 R_\perp[\eta']^2 } } \left( \frac{1}{\eta ^2 \eta'^2}-\frac{2}{\eta ^3 (\eta'-\eta )}+\frac{1}{\eta ^2 (\eta'-\eta )^2}+\frac{2}{\eta ^3
			\eta'} \right) . 
\end{align}}%
The step functions in these integrals arise from the integral over the string's length in eq. \eqref{Weyl_ScalarContribution_4DRadiation}, while the remaining 2D spatial integrals are collapsed by the Dirac delta functions in eq. \eqref{StressEnergyTensor_String_Ultrarelativistic}. Viewed as a $2+1$ dimensional problem, these step functions tell us to sum over the contributions from inside the past acoustic cone of the observer at $(\eta,\vec{x}_\perp)$. Since our string is supersonic, however, that means integrating over the contributions from its spacelike trajectory with respect to the 3D version of the fictitious Minkowski spacetime of eq. \eqref{FakeMinkowski} -- the top right panel of Fig. \eqref{CherenkovFigure_SpacetimePerspective} then reminds us, that amounts to integrating from the earlier to the later time of intersection with the backward acoustic cone; namely, $\eta_r^-$ to $\eta_r^+$, where
\begin{align}
\eta - \eta_r^\pm 	&= \sqrt{3} R_\perp[\eta_r^\pm] , \label{AcousticCone_CosmicString} \\
\eta_r^\pm 			&=  \frac{ 3 w^1 - \eta }{ 2 } \pm \frac{\sqrt 3}{ 2 } \sqrt{ ( w^1 - \eta)^2 - 2 (w^2)^2 }  .
\end{align}
Moreover, the argument of the square root in the denominator of eqs.~\eqref{I1}-\eqref{I4} is in fact the square of the geodesic acoustic distance between $(\eta,\vec{x}_\perp)$ and $(\eta',\eta'\widehat{p}_0)$ within the fictitious Minkowski spacetime \eqref{FakeMinkowski}, which factorizes as
\begin{align}
	(\eta-\eta')^2 - 3 R_\perp[\eta']^2
	= 2 (\eta_r^+ - \eta') (\eta' - \eta_r^-) .
\end{align}
Finally, in equations \eqref{I3} and \eqref{I4} we have performed a partial fractions decomposition, to demonstrate the close relations between equations \eqref{I1} through \eqref{I4}. Taking into account the discussions in the present and previous two paragraphs, we see that equations \eqref{I1} through \eqref{I4} can all be derived from the single master integral
{\allowdisplaybreaks\begin{align}
	\label{MasterIntegral_MainText}
	I[\eta;\eta_r^-,\eta_r^+] 
	&\equiv \int_{\eta_r^-}^{\eta_r^+} \frac{\dd \eta'}{(\eta - \eta')\sqrt{(\eta_r^+ - \eta')(\eta' - \eta_r^-)}} \nonumber\\
	&= \frac{\pi}{\sqrt{(\eta-\eta_r^+) (\eta-\eta_r^-)}} \qquad \text{(if $\eta > \eta_r^+ > \eta_r^-$)} \\
	&= -\frac{\pi}{\sqrt{(\eta-\eta_r^+) (\eta-\eta_r^-)}} \qquad \text{(if $\eta < \eta_r^- < \eta_r^+$)} ; \nonumber
\end{align}}%
where all the square roots are positive ones. Some integrals, for e.g. eq. \eqref{I2}, requires differentiating $I[\eta;\eta_r^-,\eta_r^+]$ once with respect to $\eta$. Others like eq. \eqref{I1} require differentiation with respect to $\eta$ followed by setting it to zero. We evaluate eq. \eqref{MasterIntegral_MainText} in \S \eqref{Section_Integral} below. For now, we will simply allow eq. \eqref{MasterIntegral_MainText} to lead us from eq. \eqref{ScalarTidalForces_String_Relativistic} to
{\allowdisplaybreaks
\begin{align}
\label{ScalarTidalForces_String_Relativistic2} 
\delta_1  C^{(\Psi|\text{Direct})i}{}_{0j0}[\eta,\vec x] 
&= - \frac{ 8 G_\text N \mu p_0}{a[\eta]^2} \Theta\left[ \eta - w^1 - \sqrt{2} |w^2| \right] \nonumber\\
&\times  \bigg(  \left( \delta_{ij} -  3 \widehat n^i \widehat n^j \right) D_1[\eta,\vec x_\perp] - \vph  \left( \widehat p_0^i \widehat p_0^j -  \widehat n^i \widehat n^j \right) D_2[\eta,\vec x_\perp] \\
&- \left( \widehat p_0^i (\widehat{n} \times \widehat{p}_0)^j + \widehat p_0^j (\widehat{n} \times \widehat{p}_0)^i \right) D_3[\eta,\vec x_\perp]  
-  \left( (\widehat{n} \times \widehat{p}_0)^i (\widehat{n} \times \widehat{p}_0)^j - \widehat n^i \widehat n^j \right)  D_4[\eta,\vec x_\perp]  \bigg) ,  \nonumber
\end{align}}
where
{\allowdisplaybreaks
\begin{align}
	D_1[\eta,\vec x_\perp] &= \frac{\pi}{2 \sqrt 2} \frac{  \eta_r^+ + \eta_r^- }{ \left( \eta_r^+  \eta_r^-\right)^\frac32}  ,  \\
    D_2[\eta,\vec x_\perp] &= \frac{ 9 \pi}{2 \sqrt 2} \frac1{\eta^3} \Bigg(    \frac{ ( w^1 - \eta ) \left(   4 \sqrt 3 R_\perp[\eta_r^+] R_\perp[\eta_r^-] w^1  + \left( R_\perp[\eta_r^+] + R_\perp[\eta_r^-] \right)  \eta ( w^1 - \eta )  \right)  }{ 3 \left(  R_\perp [\eta_r^+] R_\perp [\eta_r^-] \right)^\frac32 }  \nonumber\\
	&\qquad\qquad\qquad\qquad\qquad\qquad
	+  \frac{ \left( \eta_r^+ + \eta_r^- \right) (w^1)^2 \eta + 4 \eta_r^+ \eta_r^- w^1 (w^1 - \eta) }{ \left( \eta_r^+  \eta_r^-\right)^\frac32}  \Bigg)   ,  \\
	D_3[\eta,\vec x_\perp] 
	&= \frac{ 9 \pi}{2 \sqrt 2} \frac {w^2}{\eta^3} \Bigg(    \frac{   2 \sqrt 3 R_\perp[\eta_r^+] R_\perp[\eta_r^-] ( 2 w^1 - \eta )  + \left( R_\perp[\eta_r^+] + R_\perp[\eta_r^-] \right)  \eta ( w^1 - \eta )    }{ 3 \left(  R_\perp [\eta_r^+] R_\perp [\eta_r^-] \right)^\frac32 }  \nonumber\\
	&\qquad\qquad\qquad\qquad\qquad\qquad+  \frac{ \left( \eta_r^+ + \eta_r^- \right) w^1 \eta + 2 \eta_r^+ \eta_r^-  ( 2 w^1 - \eta) }{ \left( \eta_r^+  \eta_r^-\right)^\frac32}  \Bigg)  ,  \\
	D_4[\eta,\vec x_\perp]  & =  \frac{ 9 \pi}{2 \sqrt 2} \frac{ ( w^2 )^2}{\eta^3} \left(    \frac{   4 \sqrt 3 R_\perp[\eta_r^+] R_\perp[\eta_r^-]  + \left( R_\perp[\eta_r^+] + R_\perp[\eta_r^-] \right)  \eta    }{ 3 \left(  R_\perp [\eta_r^+] R_\perp [\eta_r^-] \right)^\frac32 }  +  \frac{ \left( \eta_r^+ + \eta_r^- \right)  \eta + 4 \eta_r^+ \eta_r^-  }{ \left( \eta_r^+  \eta_r^-\right)^\frac32}  \right)  .
\end{align}}%
Here, we remind the readers that the conformal spatial distance $R_\perp [\eta_r^\pm]$ on the perpendicular 2D plane, defined in eq.~\eqref{2DConformalDistance}, can be replaced with $(\eta-\eta_r^\pm)/\sqrt{3}$, as per eq.~\eqref{AcousticCone_CosmicString}.

{\bf Tidal Forces Near the Cherenkov Wedge} \qquad As the observer approaches the Cherenkov wedge, namely,
\begin{align}
	\sqrt 2 \, |w^2| \approx \eta - w^1 , \qquad \eta_r^\pm \approx \frac{  3 w^1 -  \eta }2, 
	\label{CherenkovWedge}
\end{align}
these $D$'s then become approximately
{\allowdisplaybreaks
	\begin{align}
		D_1[\eta,\vec x_\perp] & \approx \frac{2 \sqrt{2} \pi}{ (3w^1 - \eta)^2 } , \\
		D_2[\eta,\vec x_\perp] & \approx \frac{2 \sqrt{2} \pi}{  (3w^1 - \eta)^2 } , \\
		D_3[\eta,\vec x_\perp] & \approx \text{sgn}[w^2]\frac{ 4 \pi}{  (3w^1 - \eta)^2 } , \\
		D_4[\eta,\vec x_\perp] & \approx \frac{ 4 \sqrt{2} \pi}{  (3w^1 - \eta)^2 } .
\end{align}}%
Unlike the point particle case, the scalar tidal forces sourced by the cosmic string remain finite near the Cherenkov wedge:
{\allowdisplaybreaks
\begin{align}
\label{Weyl_String_UC_NearWedge}
\delta_1  C^{(\Psi|\text{Direct})i}{}_{0j0}[\eta,\vec x] 
&\approx - \frac{ 16 \sqrt 2 \pi G_\text N \mu p_0}{a[\eta]^2 (3 w^1 - \eta)^2 } \Theta\left[ \eta - w^1 - \sqrt{2} |w^2| \right] \left(   \delta_{ij}   -  3 \widehat u^i \widehat u^j     \right) , 
\end{align}}%
with $\widehat{u}$ being the unit spatial vector normal to the Cherenkov wedge, pointing towards the region of no signal:
\begin{align}
\widehat u \equiv \frac1{\sqrt 3} \widehat p_0 + \sqrt{\frac{ 2 }3} \text{sgn}[w^2] (\widehat{n} \times \widehat{p}_0) . 
\label{UnitNormal_CS}
\end{align}

\section{Discussion and Future Directions}
\label{Section_Summary}

In this paper, we have coupled a point mass $m$ and a Nambu-Goto string with tension $\mu$ only to the gravitational field, and worked out -- via the linearized Weyl tensor in eq. \eqref{Weyl_ScalarContribution_4DRadiation} -- the Bardeen-scalar Cherenkov radiation they produce when moving at supersonic and ultra-relativistic speeds in a 4D radiation dominated universe. We have shown that a hypothetical observer initially feeling no spin$-0$ gravitational waves whatsoever will suddenly be subject to a surge of traceless-tidal-forces due to the passing Cherenkov shock wave as the point mass or cosmic string zips by. Since the relativistic fluid has been completely irrelevant in this analysis and since tidal forces are capable of exerting work on a finite size body, we may therefore identify this Cherenkov shock wave as radiation due to Bardeen-scalar gravitational waves.

Consider a pair of test masses placed very close to the Cherenkov shock front, and let $\xi^{i}$ be the vector that joins one mass to the other. Recall that the geodesic deviation equation says, $a^i$, the force per unit mass on $\xi^i$ -- and, hence, on the pair of test masses -- projected along its unit length counterpart $\widehat{\xi}^i$ is approximately given by $ \vec a \cdot \widehat \xi \approx -a^{-2} \delta_1 C^{i}_{\phantom{i}0j0} \widehat{\xi}^i \xi^j$, where we have replaced Riemann with Weyl because we have argued in \cite{Chu:2020sdn} and \cite{Chu:2019ndv} that the high frequency portion of the gravitational wave induced tidal forces in the cosmological context at hand is dominated by the latter. Hence, from equations \eqref{Weyl_BH_UC_NearCone} and \eqref{Weyl_String_UC_NearWedge}, we may learn that the force per unit mass on $\xi^i$ as the Cherenkov cone or wedge passes by, goes as 
\begin{align}
\vec a \cdot \widehat \xi
&\approx  \frac{ 16 \sqrt2 G_\text N m  p_0 |\vec \xi| }{ 3^\frac14 a[\eta]^4  (3x_\parallel - \eta )^2  \sqrt{ \eta - x_\parallel }  } \frac1{ \sqrt \ell }  \Theta\left[ \eta - x_\parallel - \sqrt{2} x_\perp \right] \left( 1 - 3 \left( \widehat{\xi} \cdot \widehat{u} \right)^2 \right) + \mathcal O \left[ \ell^\frac12 \right] \label{TidalForceCherekovFront_BH}
\end{align}
for the black hole; and
\begin{align}
\vec a \cdot \widehat \xi
&\approx  \frac{ 16 \sqrt 2 \pi G_\text N \mu p_0 |\vec \xi| }{a[\eta]^4 (3 w^1 - \eta)^2 } \Theta\left[ \eta - w^1 - \sqrt{2} |w^2| \right] \left( 1 - 3 \left( \widehat{\xi} \cdot \widehat{u} \right)^2 \right)
\label{TidalForceCherekovFront_CS}
\end{align}
for the cosmic string. 

When the pair of masses are parallel or anti-parallel to the Cherenkov shock front's normal, the magnitude of the tidal forces are maximum but $\vec{a}\cdot\widehat{\xi}$ itself is negative due to the angular factors in equations \eqref{TidalForceCherekovFront_BH} and \eqref{TidalForceCherekovFront_CS}. The pair of masses are, therefore, {\it compressed}. This Bardeen-scalar tidal force reduces in strength as the orientation of the pair is rotated, $1/\sqrt{3} < |\widehat{\xi} \cdot \widehat{u}| < 1$, and reaches zero at $\widehat{\xi} \cdot \widehat{u} = 1/\sqrt{3}$. (This $3$ is related to the dimension of space -- because Weyl is traceless -- and is not related to the $\sqrt{w}$.) After that, the force becomes positive between $0 \leq |\widehat{\xi} \cdot \widehat{u}| < 1/\sqrt{3}$ and {\it stretches} them apart. Moreover, a novel feature near the Cherenkov front can be spotted in the results of equations \eqref{TidalForceCherekovFront_BH} and \eqref{TidalForceCherekovFront_CS}: for a fixed time $\eta$, the further away the observer is from the present location of the black hole or cosmic string, i.e., the earlier the retarded time $\eta_r = ( 3x_\parallel-\eta )/2$ or $\eta_r = ( 3w^1 - \eta )/2$, the stronger the scalar tidal forces. This turns out to be a purely cosmological effect, because the appearances of these retarded times in eqs.~\eqref{TidalForceCherekovFront_BH} and \eqref{TidalForceCherekovFront_CS} can be traced back to the $1/a[\eta]^2$ factors within the sources' ultra-relativistic stress tensors \eqref{PointMass_StressTensor_Ultrarelativistic} and \eqref{StressEnergyTensor_String_Ultrarelativistic}, and as a result, an earlier retarded time would correspond to a stronger source ${}^{\text{(a)}}T_{\mu\nu}$, and therefore a stronger tidal effect.

While the possibility of Cherenkov gravitational radiation has been noted previously (see, for e.g., \cite{Tolish:2016ggo}) we believe our work is the first to compute it explicity. On the other hand, we have neglected the dynamics of the background relativistic fluid thus far. It would be of physical importance to quantify the impact on the fluid due to the Cherenkov processes we are examining here, so as to properly understand if there are any observables that would remain accessible to us in the present era. Additionally, while we have studied an infinite straight cosmic string primarily for technical simplicity, more realistic string dynamics should be considered. Of particular relevance to the phenomenon of Cherenkov radiation are cusps, generic features on cosmic strings which move at the speed of light in vacuum. Other features, such as kinks, could also generate Bardeen-scalar radiation whose signatures have remained unexplored to date.

\section{Acknowledgements}

YZC was funded in part by MOST 109-2112-M-008-017. YWL was supported by the Ministry of Science and Technology of the R.O.C. under Project No.~MOST 109-2811-M-007-514. He also wishes to thank Kin-Wang Ng and Chong-Sun Chu for discussions.

\appendix

\section{A Cosmic String Integral}
\label{Section_Integral}

In this section we seek to evaluate
\begin{align}
	\label{MasterIntegral_Appendix}
	I[\eta;a,b] \equiv \int_{a}^{b} \frac{\dd x}{(\eta-x)\sqrt{(b-x)(x-a)}} ,
\end{align}
where the $\sqrt{\cdot}$ is the positive one, and $b>a>0$. The parameter can either be less than both $a$ and $b$ (i.e., $b > a > \eta$); or greater than both ($\eta > b > a$).

\begin{figure}[!ht]
	\begin{center}
		\includegraphics[width=2.5in]{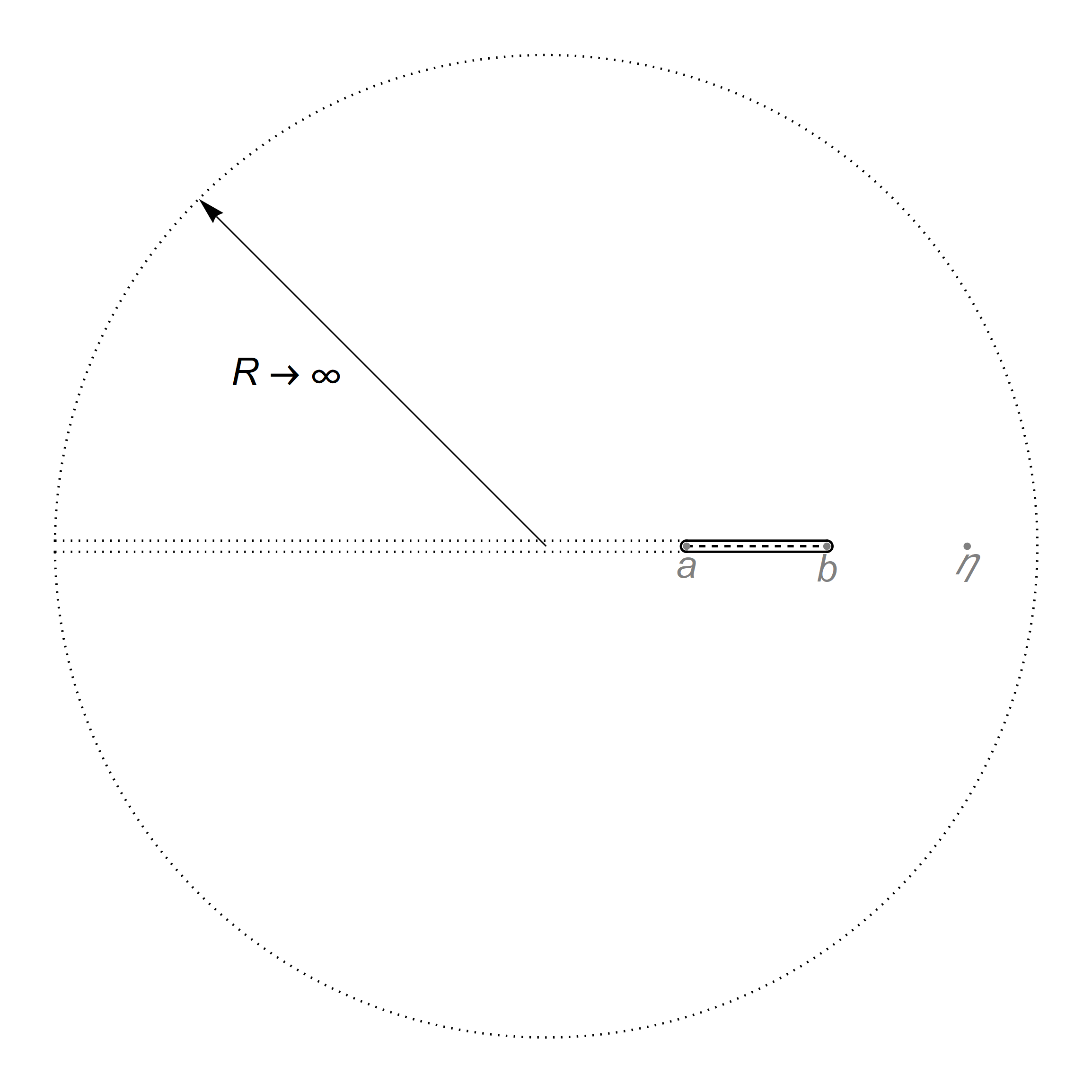}
		\includegraphics[width=2.5in]{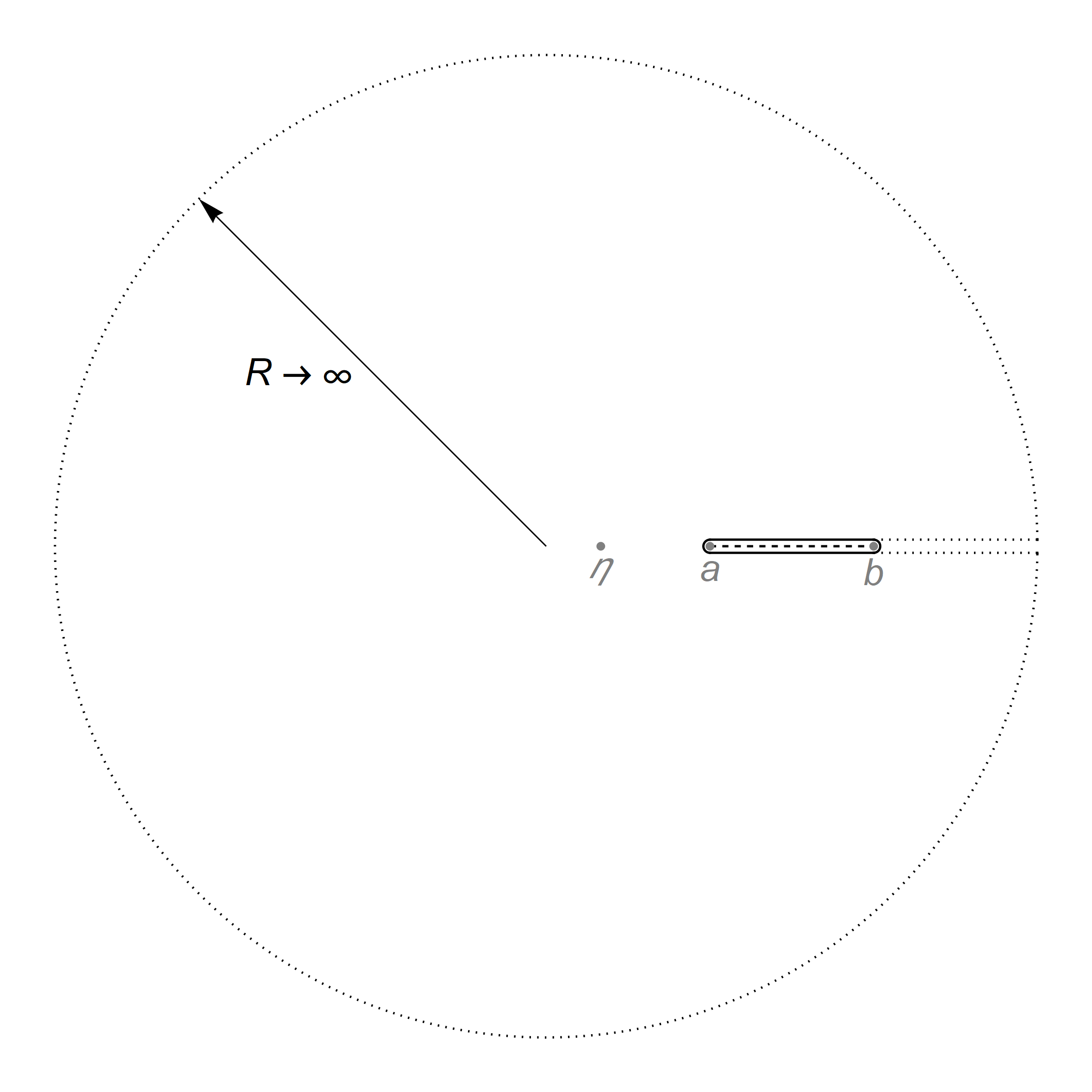}
		\caption{{\it Cosmic string integral:} Both panels represent the complex $z$ plane. The $a$, $b$, and $\eta$ lie on the Real line. We choose the branch cut to be the black dashed line joining $a$ and $b$. The $C_1+C_2$ contour in the main text refers to the clockwise closed loop in black encircling $a$ and $b$. {\it Left Panel:} When $\eta > b > a$, we may extend the $a$-end of this closed loop off to $-\infty$ on the Real line, and then make a radius $R \to \infty$ counterclockwise circle back to itself. This converts the small black clockwise closed loop into the large counterclockwise dotted closed loop, whose associated integral can be evaluated via the residue at $z=\eta$. {\it Right Panel:} On the other hand, when $\eta < a < b$, we may extend the $b$-end of the black closed loop off to $+\infty$ on the Real line, and then make a radius $R \to \infty$ counterclockwise circle back to itself. The integral around the small black clockwise closed loop is therefore equal to that around large counterclockwise dotted closed loop, which in turn can be evaluated through the residue at $z=\eta$.}
		\label{ContourIntegrals}
	\end{center}
\end{figure}

Our strategy is to relate $I$ to a closed loop integral, by first identifying continuing $x \to z$ onto the complex plane as follows 
\begin{align}
	\sqrt{(b-z)(z-a)} &= \sqrt{\rho_1 \rho_2} \exp\left[ \frac{i}{2} \left( \phi_1 + \phi_2 - \pi \right) \right], \qquad
	0 \leq \phi_{1,2} < 2\pi .
\end{align}
Here, $\rho_1 = |z-a|$ and $\rho_2 = |z-b|$. The angles $\phi_{1,2}$ are, respectively, the angles $z-a$ and $z-b$ makes with the positive Real axis in a counter-clockwise manner. We will choose the branch cut to run along the line joining $a \leftrightarrow b$; so we are not allowed to cross it. Observe that, slightly above the branch cut, we have $\phi_1 = 0$ and $\phi_2 = \pi$, which in turn implies
\begin{align}
	\sqrt{(b-z)(z-a)} 
	= \sqrt{|z-a||z-b|} \exp\left[ \frac{i}{2} \left( 0 + \pi - \pi \right) \right] 
	= \sqrt{|z-a||z-b|},
\end{align}
so that the integrand coincides with the original one in $I$. When $z^\pm$ lies on the positive Real axis greater than $b$, we may check it is single-valued because we have either 
\begin{align}
	\sqrt{(b-z)(z-a)} = \sqrt{|z-a||z-b|} \exp\left[ \frac{i}{2} \left( 0 + 0 - \pi \right) \right] 
\end{align}
or
\begin{align}
	\sqrt{(b-z)(z-a)} = \sqrt{|z-a||z-b|} \exp\left[ \frac{i}{2} \left( 2\pi + 2\pi - \pi \right) \right] .
\end{align}

Let us now consider the closed contour $C_1 + C_2$, where $C_1$ goes along a straight line from $a \to b$ just above the branch cut; while $C_2$ goes from $b \to a$ just below it. Note that the small semi-circle of radius $ \rho_1 = \epsilon \ll 1$ around $a$ would go contribute as
\begin{align}
	\int_0^\pi \frac{\epsilon \cdot e^{i\theta_1} i \dd\theta_1}{(\eta-z) \sqrt{\rho_2 \cdot \epsilon} e^{i( \theta_1 + \theta_2 - \pi )/2}} \sim \sqrt{\epsilon} ,
\end{align}
and are therefore negligible as $\epsilon \to 0$; similar remarks apply to the small semi-circle around $b$ as well. Hence, $C_1 + C_2$ form a closed contour integral; and let us deduce it is actually twice of our original integral:
\begin{align}
	I_{C}[\eta;a,b] 
	&\equiv \int_{C_1+C_2} \frac{\dd z}{(\eta-z)\sqrt{(b-z)(z-a)}} \\
	&= I[\eta;a,b] - \int_{a}^{b} \frac{\dd x}{(\eta-x) |x-a|^{\frac{1}{2}}|x-b|^{\frac{1}{2}} e^{(i/2)(2\pi + \pi - \pi)}} \\
	&= 2 I[\eta;a,b] 
\end{align}
\underline{$\eta > b > a$} \qquad Referring to Fig. \eqref{ContourIntegrals}, for this first case let us extend the closed contour $C_1 + C_2$ by running one line just above and another just below the Real axis -- but in opposite directions, so their contributions exactly cancel -- from $a$ to $-\infty$. Then draw a infinitely large counter-clockwise circle $C_\infty$ centered at $0$ that touches these two lines at $-\infty$. This infinitely large circle would contribute as
\begin{align}
	\lim_{R \to \infty} \left\vert \int_{C_\infty} \frac{\dd z}{(\eta-z)\sqrt{(b-z)(z-a)}} \right\vert
	\leq \lim_{R \to \infty} \int_{0}^{2\pi} \frac{R \dd \theta}{R^2} \to 0 .
\end{align}
We may invoke the Residue Theorem and compute
\begin{align}
	I[\eta;a,b] 
	&= \frac{1}{2} \oint_{C_1+C_2+C_\infty} \frac{\dd z}{(\eta-z)\sqrt{(b-z)(z-a)}} \\
	&= \frac{\pi}{\sqrt{(\eta-a)(\eta-b)}} ,
\end{align}
where $\sqrt{(\eta-a)(\eta-b)} > 0$.

\underline{$b > a > \eta$} \qquad For this second case, still referring to Fig. \eqref{ContourIntegrals}, let us extend the closed contour $C_1 + C_2$ by running one line just above and another just below the Real axis -- but in opposite directions, so their contributions exactly cancel -- from $b$ to $+\infty$. Then draw a infinitely large counter-clockwise circle $C_\infty$ centered at $0$ that touches these two lines at $+\infty$. This infinitely large circle would contribute $0$ as before. Then, we again employ the Residue Theorem to deduce
\begin{align}
	I[\eta;a,b] 
	&= \frac{1}{2} \oint_{C_1+C_2+C_\infty} \frac{\dd z}{(\eta-z)\sqrt{(b-z)(z-a)}} \\
	&= - \frac{\pi}{\sqrt{(a-\eta)(b-\eta)}} ,
\end{align}
where $\sqrt{(a-\eta)(b-\eta)} > 0$.

\end{document}